# Field Effect Transistor Nanosensor for Breast Cancer Diagnostics

Pritiraj Mohanty[a], Yu Chen[a], Xihua Wang[a], Mi K. Hong[a], Carol L. Rosenberg[b], David T. Weaver[c], Shyamsunder Erramilli[a]

[a]Department of Physics, Boston University, 590 Commonwealth Avenue, Boston, MA 02215

[b]Boston University School of Medicine, Boston, MA 02215

[c]Ninth Sense, Inc., 8 Saint Mary's Street, Boston, MA 02215

## 1. Abstract

Silicon nanochannel field effect transistor (FET) biosensors are one of the most promising technologies in the development of highly sensitive and label-free analyte detection for cancer diagnostics. With their exceptional electrical properties and small dimensions, silicon nanochannels are ideally suited for extraordinarily high sensitivity. In fact, the high surface-to-volume ratios of these systems make single molecule detection possible. Further, FET biosensors offer the benefits of high speed, low cost, and high yield manufacturing, without sacrificing the sensitivity typical for traditional optical methods in diagnostics. Top down manufacturing methods leverage advantages in Complementary Metal Oxide Semiconductor (CMOS) technologies, making richly multiplexed sensor arrays a reality. Here, we discuss the fabrication and use of silicon nanochannel FET devices as biosensors for breast cancer diagnosis and monitoring.

## 2. Introduction

A biosensor, described succinctly by Ziegler and Gopel(Ziegler & Gopel, 1998), is "a device integrated with a biological sensing element, which is a product derived from a living system and a signal transducer which can provide a recognition signal of the presence of a specific substance"(Hall, 1991; Ziegler & Gopel, 1998). The biological sensing element, which determines the specificity of the biosensor, can be composed of an enzyme, an antibody, a nucleic acid, or another analyte detecting molecule. The specific binding or reaction between the target and the receptor (or the biological sensing element) introduces a signal that is then transduced and measured. Because of these fundamental properties, biosensors can be configured for macromolecular recognition, such as with human cells of different types, viruses, and pathogenic organisms. Therefore, there is a far-reaching diagnostic utility in these devices in applications ranging from human health to food safety, drug response, and personalized medicine.

Biosensors may be categorized by the operational mechanism of the sensors(Byfield & Abuknesha, 1994). Although optical biosensors using colorimetric, fluorescence, luminescence, and absorbance are industry and diagnostics standards, these strategies necessitate target labeling and amplification. Also, the instrumentation footprint necessary to sensitively read optics-based signals is large compared with that achievable in devices incorporating nanotechnologies and microelectronics. Thus, emerging technologies that improve the sensitivity, cost, instrumentation, or field applicability of biosensors are beginning to be implemented. Mechanical biosensors utilize mass loading during the recognition process, introducing a dynamic resonance frequency change or static deflection of the device. Electronic biosensors measure the change of the capacitance or conductance of the device due to biological recognition.

## 2.1. Nanoelectrical Sensors

A field effect transistor (FET) uses an electric field to control the electrical channel of conduction, and hence the conductivity of the charge carriers in the channel. The flow of charge carriers between the source and the drain can be tuned by modifying the size and the shape of the conducting channel by applying an electric field to the gate. In the biosensor configuration, the FET consists of a nanowire channel between the source and the drain terminals. The nanowire surface can be bio-functionalized so that a biomolecular binding event can create an electric field, similar to the control electric field applied to a conventional FET (Figure 1). In devices that use the FET principle, a designated, physically separated sensor surface is formed by precision manufacturing. The FET sensor is connected to an electronic circuit to monitor the specific conductance of this sensor surface.

FET biosensors are adapted for the measurement of biomolecules interacting with such a sensor surface (Figure 1). As with other forms of sensing, the surface of a FET biosensor is modified to selectively recognize specific analytes. In the illustration, antibodies are conjugated to the surface as part of the manufacturing process. These antibodies are selected for specific detection of a protein of interest. Molecular binding events between the analyte and the antibodies on the biosensor surface cause changes to the biosensor surface charge density and/or surface potential. In this manner, precision manufacturing of FET biosensors allows sensitive analyte recognition. The differential conductance amplitude is correlated to the analyte concentration in the sample solution.

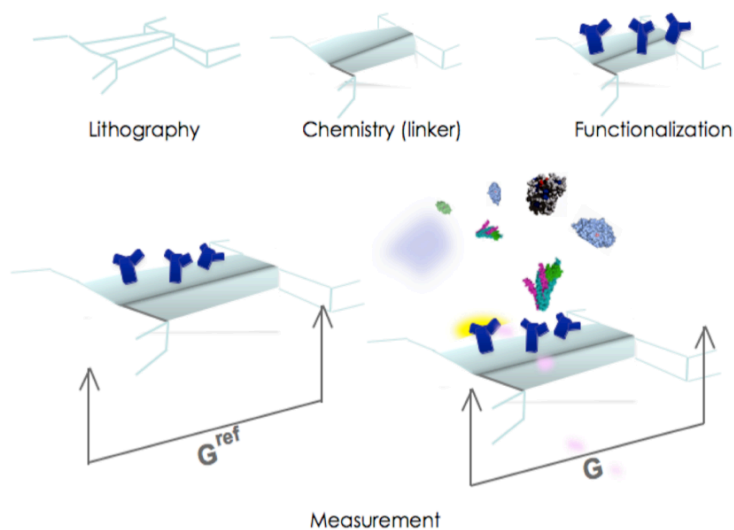

Figure 1. FET biosensor manufacturing and measurement. The diagram indicates the stepwise manufacturing of FET devices for diagnostic tests. Highly sensitive nanosensors are formed with precise dimensionality and surface area. Following a process of lithography and chemistry, antibodies are conjugated to the surface of the sensor (blue). Analyte measurements are conducted with samples containing heterogeneous mixtures of disease-relevant proteins, such as those that occur in blood, saliva, and other fluids. Specific analyte binding contributes to a surface charge differential, detected electrically as a change in conductance ($\Delta G$) across the nanosensor surface.

The field effect transistor was first introduced as a sensor for analyte ion concentration measurements in 1970 (Bergveld, 1970), and was later adapted for many different sensor applications (Chung *et al.*, 2006; Hammond & Cumming, 2005). In 1997, detection of DNA hybridization was reported using a FET sensor (Souteyrand *et al.*, 1997). When these technologies were introduced, the operational dimensions of the device were in the micron-millimeter range, with recording buffer concentrations on the order of a millimole. Others described DNA hybridization detection with a FET sensor that monitored nanomolar DNA concentrations, but a similar operational area was still required (Fritz *et al.*, 2002).

Important innovations have now been introduced which allow these measurements to be taken to the nanoscale. Although the molecular binding events are identical to those in larger devices, the nanoscale has the advantage of increased sensitivity of detection. In their pioneering work, Lieber and co-workers used chemically-grown silicon nanowire FETs for sensing (Cui *et al.*, 2001b). These nanochannel FET sensors demonstrated significant advantages of real-time, label-free and highly sensitive detection of a wide range of analytes, including proteins (Wang *et al.*, 2005), nucleic acids (Hahm & Lieber, 2004), small molecules (Chen *et al.*, 2008; Wang *et al.*, 2005; Wang *et al.*, 2008), and viruses (Patolsky *et al.*, 2004) in single-element or multiplex format (Zheng *et al.*, 2005).

A significant breakthrough has been achieved in the utilization of CMOS-compatible materials, such as silicon, to develop FET biosensors. Common nanoscale architectures include nanowires, nanotubes, nanopores, micro- or nanocantilevers, and nanoparticles (including quantum dots). Together with carbon nanotubes, silicon nanowires are considered to be the fundamental building blocks for scaling microelectronics down to the nano level. However, current carbon nanotube synthetic methods produce mixtures of metallic and semiconductor nanotubes, indicating the difficulty of fabricating scalable, homogeneous devices reproducibly.

Silicon technology is one of the most developed and well-studied technologies for the microelectronics industry. Therefore, many manufacturing, miniaturization, and multiplexing capabilities are advantageously adaptable to silicon-based nanosensors. Further, the doping of silicon can be well controlled, ensuring that silicon nanowires can be consistently made semiconducting, whether synthesized by chemical methods (chemical vapor deposition or laser ablation) or fabricated by a lithography process. Since silicon nanowires are configurable to ultra-small dimensions (20 nm in diameter and 2 μm in length), these devices promise ultra-high sensitivities sufficient to measure a single bonding event (such as a single virus particle detection).

There are several technical improvements which illustrate the importance of size effects and justify the development of nanoscale FET biosensors. First, biological recognition at the sensor surface is reflective of the entirety of the physical signal across the whole sensor device. Thus, the sensitivity of the device, relating to its signal-to-noise ratio, is improved by increasing the surface-to-volume ratio. Whereas attachment of a biomolecule to a macroscale planar device affects the "surface" characteristics, attachment to a nanoscale wire affects the "bulk" characteristics. Second, many of the biological molecules of interest in biomedical, environmental, and food applications are known to have sizes on the micro- or nanoscale. Thus, a device with submicron or nanoscale dimensions will have the best efficiency as a biological sensor. Third, as the device size is reduced to the nanoscale, properties such as the mobility of the charge carrier change (Elfstrom *et al.*, 2007) finite size effects dominate the device character, and nanodevice performance is enhanced.

Nanowires have many applications, including sensitivity enhancement for surface plasmon resonance biosensors (Byun *et al.*, 2005; Byun *et al.*, 2007; Kim, 2006). Surface-modified silicon nanochannels bind to charged groups and ions and serve as nanoscale pH sensors (Chen *et al.*, 2006). Nanowires composed of conducting polymers (polyaniline) can be used to coat the electrode for voltammetric detection of DNA hybridization (Chang *et al.*, 2007; Zhu *et al.*, 2006) as well as for the detection of the food-borne pathogen, Bacillus cereus(Kim *et al.*, 2006). Nanotechnologies are driving the development of low cost, portable FET devices for clinical applications that will greatly improve diagnostic efficiency. The promise of the nanoscale semiconductor technology lies in its potential for highly parallel detection of thousands of target molecules in tandem, and at very low concentrations relevant to their normal and diseased state concentrations in blood and other human bodily fluids. Therein lies a fundamental advantage over traditional screening methods that ordinarily detect one type of protein or biomarker per assay. Screening and diagnosis by conventional techniques does not allow for the check of a large number of biomarkers due to technological and cost limitations. Array chip techniques for both proteomics and genomics often require several stages of amplification for reliable analysis. The time and the number of cells needed for these techniques are proving to be bottlenecks for the future molecular applications in medical science. The nanosensor technologies described here are designed to overcome these challenges, as they require short analysis time with small sample volume.

Perhaps the most promising application is the use of these sensors for monitoring and detecting specific molecules associated with a particular disease, such as in cancer patient diagnostics. The FET nanosensor may logically be used for early-stage detection of cancer, a strategy crucial to the prevention of cancer-related death.

## 2.2. Breast cancer diagnostics

Human breast cancer is a clonal disease, presumed to develop when a cell acquires sufficient germline or somatic abnormalities to be transformed and to express full malignant potential (Balmain *et al.*, 2003). Although originally considered to be a single disease of the breast, extensive research and clinical studies have led to the realization that breast tumors are biologically heterogeneous. Hence, an individual patient is known to require therapies that are tailored to her specific cancer. The development of microarray RNA techniques, combined with a search for appropriate biomarkers, has led to the realization that associated with each breast cancer there may be changes in the expression level in hundreds of genes (Branca, 2003; Van den Eynden *et al.*, 2004). Traditional medical treatment protocols are unable to properly consider tumor heterogeneity, and thus may be prescribing treatments that may not be effective for a particular patient (Garber, 2004; Ring & Kroetz, 2002).

Currently, the major biological distinctions between individual tumors are made by evaluating expression of estrogen receptor (ER), progesterone receptor (PR) and Her2/*neu*, the product of the c-erbB-2 oncogene. Classification of tumors using a set of biomarkers (Weigelt *et al.*, 2003) suggests that each patient's tumor may be associated with a unique "fingerprint". Individually or in clusters, these biomarkers may have prognostic value (i.e., estimating outcome) and/or predictive value (i.e., estimating response to specific treatments, or in predicting survival in metastasized breast cancers (Bild *et al.*, 2006; Minn *et al.*, 2005; Weigelt *et al.*, 2003). This new understanding has the potential to lead to truly patient-specific treatment protocols (Garber, 2004), but it must be supported by accurate, immediate, and comprehensive evaluation of biomarkers by new diagnostics strategies.

## 3. Materials

## 3.1. Wafer-level fabrication in silicon-based materials

*Silicon wafers*: Silicon nanowires are fabricated from standard SOI (Silicon-On-Insulator) wafers using nanolithography and surface micromachining. The starting wafer has a typical silicon thickness of 30-100 nm on top of a sacrificial oxide layer varying in thickness from 100-500 nm. SOI wafers are used to replace the traditional silicon wafers in order to reduce parasitic device capacitance and thus improve performance of the devices in microelectronics (Celler & Cristoloveanu, 2003). The wafer used in our experiment is an SOI wafer (100) (University Wafer), which includes a device layer (100 nm thick), an insulation layer (380 nm thick) and a boron-doped substrate (around 600 μm thick). The volume resistivity of the device layer is 10-20 ohm·cm for these experiments.

*Design features*: We refer to the manufactured surface consisting of 1 or more silicon nanowires as a 'chip'. Typically, the nanosensors are designed with 10-100 nanowires per chip. Silicon nanowires are fabricated by exposure on three sides to provide three-dimensional relief. The associated benefit to this geometry is an increase in surface-to-volume ratio of the silicon nanowire, which enhances the dielectric effect of molecular binding. Second, we vary the system size to determine optimal structure size. Both the silicon nanowire cross-sectional dimensions and the proximity of the gates are changed in a series of devices. The main manufacturing variance parameter to tune the sensitivity range is the dimensional width and length of the nanowire between the source and drain. The nanosensor thickness, which will be nearly identical for all wires on a single chip, is chosen to be 30-100 nm in order to maintain high sensitivity, nanosensor reproducibility, and stability. The bias voltage applied on each nanosensor is also a contributor to modulating sensitivity.

We also fabricate nanowire sensors with multiple gates. Additional gates enable better sensitivity by providing a means to fine tune the nanowire conductance channel. In addition, we fabricate arrays of nanowires for differential measurement in anticipation of diagnostic need. In the differential configuration, the detected signal is the difference between the signal from a functionalized nanowire and an un-functionalized nanowire. This technique allows a more accurate determination of the biomolecular signal.

### 3.2. Nanochannel surface chemistry and functionalization

The silicon-based nanochannels contain an oxide layer (e.g., $Al_2O_3$ grown by atomic layer deposition) added during the fabrication process. This layer can be used for functionalization of the analyte detection molecules, which may be proteins, receptors, antibodies, nucleic acids, ligands, or small molecules, using chemical modification protocols extensively published in the literature.

*a) Silane treatment*

The silicon oxide surface is modified using 3-aminopropyltriethoxysilane (APTES, Sigma-Aldrich). The amine group of APTES reacts with glutaraldehyde to provide an aldehyde group that can then form an imine linkage to the primary amine group on the detector receptor proteins. Other ligands, polymers, and chemicals may be conjugated to the surface by the same or related chemistries. Alternatively, the surface of the silicon nanowire is exposed to an ethanol solution of 3-(trimethodxysilyl)propyl aldehyde to yield an aldehyde-terminated nanosensor surface. Silanisation is conducted after the Al2O3 coating steps. The nanowire device is placed into a hot water bath for a short time to clean the surfaces. The device is then blown dry with N2 gas. The device is silanised by 5% APTES (3-aminopropyltriethoxysilane, Gelest, Inc.) solution in 95% ethanol, 5% DI water for two hours. APTES is removed by rinsing with ethanol (>3 times) and DI water (>3 times) and then dried with N2 gas following a bake at 120° C for 30 minutes. The conjugation-ready receptors (section below) are then directly coupled to these aldehyde surfaces.

*b) Antibody conjugation*

Immobilizing a selected antibody or receptor to the modified surface contributes the desired specificity for molecular interactions to be studied. These molecules are prepared for linking to the aldehyde surfaces by crosslinking chemistry. First, EDC and sulfo-NHS (Sigma-Aldrich) are used to crosslink to the detector protein by standard conjugation protocols. For the studies reported here, the biotinylated monoclonal CA15.3 from Fitzgerald (host species – mouse) recognizing the CA15.3 tumor antigen is used. Second, after the aldehyde linked surface has been bound to the antibodies to CA15.3 or other analyte detections of interest, it is important to make sure that there are no free (i.e., unreacted) aldehyde surface groups that remain. A potential artifact can result from non-specific binding of other proteins and peptides to the unbound aldehyde links. Passivation of the surface with ethanolamine is sufficient, and it is routinely checked using fluorescently labeled peptide markers. Functionalization is optimized and checked by using an immunoglobulin assay (human IgG). The same procedure is operational for other molecules with primary amine groups.

## 4. Methods

The invention of the transistor was a defining point for the modern electronics revolution. An accelerator of this technology occurred with the realization of techniques to shrink the physical dimensions of the transistors to the nanoscale. This capability to manufacture at the nanoscale is also expected to facilitate an important breakthrough in diagnostics. Fabrication of the nanowire sensor involves a series of processes, including pattern definition by electron-beam lithography, material deposition by thermal evaporation or e-beam evaporation, plasma cleaning, lift-off and a number of dry (reactive-ion plasma) and wet (acid) etch processes. We have succeeded in developing manufacturing processes for a number of materials, including insulating GaAs, semi-insulating or doped silicon, silicon oxide, and metallic suspended structures.

### 4.1. Nanochannel device fabrication

*a) Photolithography*

Photolithography is currently the most widely used fabrication method in industrial settings for submicron or nanoscale structures. During the stepwise photolithography process, a mask is used to define the structure on the wafer. The wafer is first spin-coated with a polymer (photoresist) that is sensitive to the UV light. After soft baking, the wafer is placed beneath the photomask and intensely exposed to the UV light. The more common type of resist, positive photoresist, becomes more soluble in developer after exposure to UV light (negative photoresist, by contrast, becomes less soluble). The developer is then applied to remove the photoresist pattern that has been exposed to the UV light.

The resolution of photolithography is limited by the wavelength of the UV light. In current technology, UV light with a wavelength of 183 nm is used to pattern features as small as 37 nm. Smaller wavelengths are able to create smaller structures, but as demand dictates further miniaturization, new optics which utilize reflection strategies are required. Other technologies, such as phase shifting masks or immersion of high refractive efficient media between the wafer and apertures, can be used to pattern structures smaller than 50 nm.

*b) Electron beam lithography*

Electron beam lithography, laser beam lithography, and focused ion beam lithography are serial processes used to pattern micro/nano structures. Because the throughput is low, electron beam lithography is used primarily to produce photomasks or in research laboratory settings. It is sufficient, however, to develop and test design fabrication prototypes for the nanochannel FET devices. The typical process is diagrammed in a stepwise manner (Figure 2). First, electron beam lithography is used to define source drain electrodes. Two layers of PMMA (polymethylmethacrylate) are spun on the wafer surface, and exposure to a high energy electron beam breaks the PMMA bonds, making it soluble in developer. We use a developer composed of an MIBK: IPA solution for PMMA. The resolution of electron beam lithography depends on the acceleration voltage of the electron beam, and also the secondary electron travel in the resist (PMMA). Electron beam evaporation is used for metallization. Titanium and Gold are typical electrode materials, with titanium acting as an adhesive layer. After lift-off, another lithography process is used to define a nanochannel, and chromium is used as the mask for the subsequent etching process.

*c) Reactive Ion Etching*

Reactive ion etching (RIE) is a typical method for both isotropic and anisotropic etching. Chemically reactive plasma is used to remove the material on the wafer surface. The material to be removed dictates choice of plasma chemistry. For example, to etch silicon, plasmas based upon either $CF_4$ or $SF_6$ are used. While reactive ion etching is mainly a chemical reaction, the process also depends on the power and pressure, as the ions bombard the surface and transfer kinetic energy to the wafer in a process akin to sputtering. By appropriate power and pressure adjustments, we achieve a uniform anisotropic etch and straight side walls.

Wet etching can also be used in nanochannel device fabrication. Stern *et al.*(Stern et al., 2007a) applied a wet anisotropic etching technique combined with a lithography process. They have demonstrated that high quality nanochannel FET sensors can be created using 'top-down' fabrication techniques. These devices can be used as highly sensitive, specific label-free sensors down to 100 femtomolar concentrations of antibody-based detection, as well as for utilities in real-time monitoring of the cellular immune response. In this multistep process, tetramethylammonium hydroxide is used to etch Si (111) planes around one hundred times more slowly than the other planes, such that all imperfections not aligned to the (111) plane are smoothed. In their work, the oxide of the nanobar is removed and a direct silicon-carbon bond is formed by a photochemical hydrosilation reaction. Yields of successful selective functionalization by this approach have been reported to be low (around 2%), although improvement may be expected.

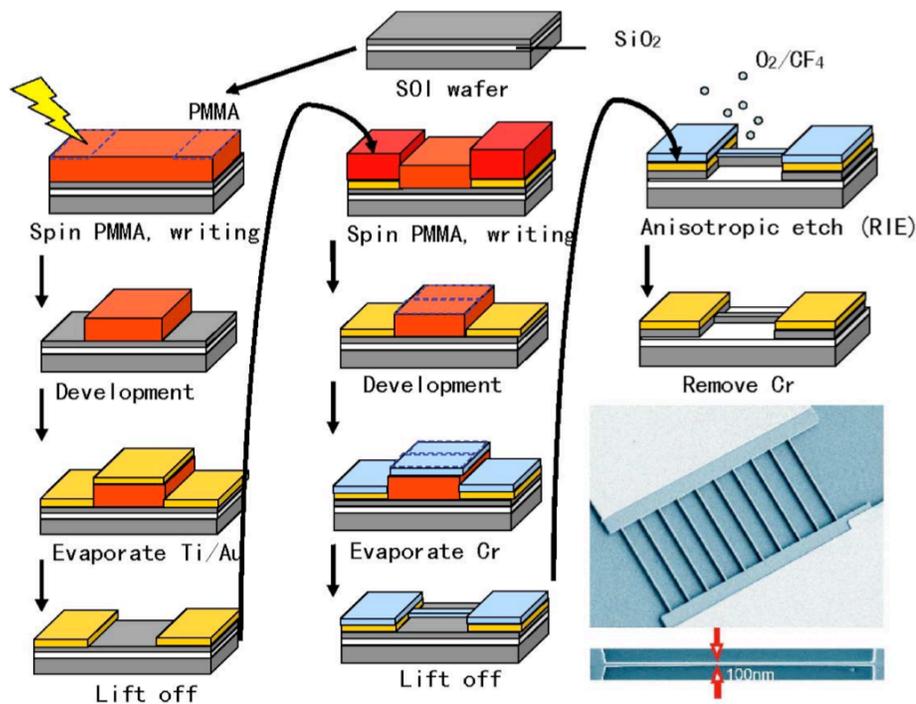

Figure 2. The stepwise fabrication process. Ti, Au and Cr denote the metals titanium, gold, and chromium, respectively. Reactive ion etching (RIE) is used to give vertical relief to the wires. Lower right photograph: scanning electron micrograph image of a suite of 10 nanosensors with critical dimensions of 100 nm, connected to a common source/drain.

Following etching, a thin layer of $Al_2O_3$ (5 to 10 nm) is deposited on the nanowire(s) by atomic layer deposition (ALD) to form a high quality insulation layer. Thin films grown by ALD are chemically bonded to the substrate and are pinhole free. The $Al_2O_3$ insulation layer has almost perfect conformity compared with other methods and can be used to deposit many types of films. A possible advantage of the $Al_2O_3$ insulation layer is that the $Al_2O_3$ surface has an increased capacity for silanization compared to silicon. $Al_2O_3$ is also used to protect the electrodes in the solution.

## 4.2. Measurement with a standard nanochannel FET device

Nanowires are particularly attractive as biosensors due to the critical dimensions of the nanostructures. The detection sensitivity is greatly enhanced as a result of the large surface-to-volume ratio achieved with nanoscale conductors, where the measured conductance is dominated by surface contributions. Therefore, the presence of charged proteins bound to the surface induces a large fractional change in the nanowire conductance, and enables relatively easy detection.

A conventional electrical circuit is used to measure the functionalized nanosensors. To establish the circuitry, of particular interest is the location of a control gate in the devices (Figure 3), which is the subject of future improvements in design, fabrication, and sensitivity. This gate has allowed us to enhance the sensitivity and to control the labeling of the nanowire, leading to a new method of using nanosensors for biomarkers. Without any gate biasing, we are able to measure the presence of the model marker at a concentration of less than 1 ng/ml. By adjusting the gate voltage, the sensitivity is increased by at least 1 order of magnitude.

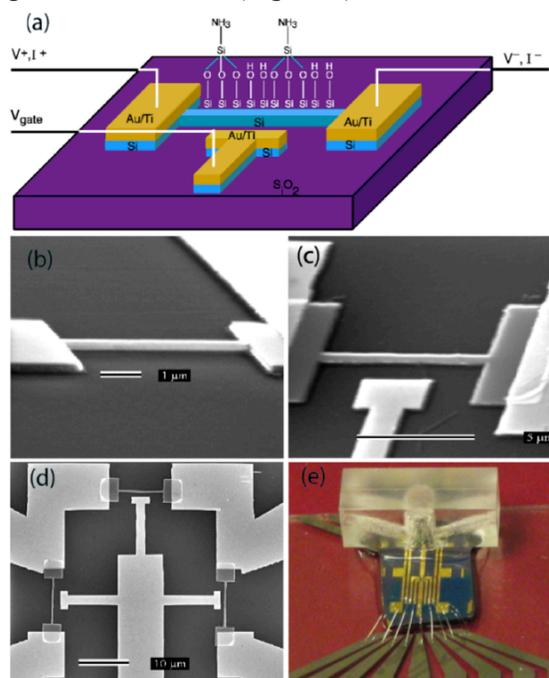

The measurement circuit includes a small AC (alternating current) modulation (provided by an EG&G 5210 lock-in amplifier), superimposed on the DC (direct current) bias across the nanowire (provided by a Keithley 2400 source meter)(Figure 3D). The AC-modulation voltage and the DC bias voltage are added by a non-inverting summing circuit integrated with the preamplifier circuit. Differential conductance measurements are taken by sweeping the DC bias at constant AC modulation amplitude and measuring the response with the lock-in amplifier, which is referenced to the AC signal frequency. The measurement of interest is $G$, the change in the differential conductance due either to a change in the reference gate voltage $\Delta V_g$, or to a change in molecular concentration $\Delta m$. PDMS gel is used to seal the device and surround the nanowire, which is bathed in a fluid volume of 20-30 microliters, connected to a syringe pump. Unlike in microfluidic cells, where laminar flow prevents complete mixing, a larger cell shortens the response time of diffusion-controlled processes.

Figure 3. Device schematic diagram, scanning electron micrographs and measurement circuit. (a) The schematic diagram of the silicon nanowire with side gates and electrodes. The nanowire is exposed on three sides along the longitudinal directions. (b) The nanowire shown here is 300 nm wide, 230 nm thick and 8μm long. (c) A silicon nanowire with a side gate. (d) The scanning electron micrograph displays three silicon nanowire devices on the same chip. (e) An optical micrograph shows the flow chamber sealed on top of the devices on the interface board.

## 5. Results

### 5.1. Breast cancer biomarkers and clinical need

Breast cancer is the most common life-threatening malignancy in women of most developed countries today, with approximately 180,000 new cases diagnosed every year. Roughly half of these newly diagnosed patients are node-negative. However 30% of these cases progress to metastatic disease. Breast cancer biomarkers have emerged as an important determinant of diagnosis, staging, progression, and therapy decision-making. The CA15.3 tumor marker is produced from the mucin 1 (MUC1) gene product and it is an epithelial marker. CA15.3 is highly overexpressed in breast, ovarian, and other cancers compared with the normal epithelium from these tissues. The protein also has the property of being cleaved from the cell surface as a soluble protein fragment, and it is therefore in the blood stream. Many studies have demonstrated that breast cancer patients have elevated CA15.3 levels.

## 5.2. Protein sensing and signal amplification – general features

A change in conductance is primarily due to the contribution of surface states to the conductance, which for larger sensors is dominated by volume effects. The fractional change is greatest for the smallest sensors, due to the increased surface-to-volume ratio.

Small changes in the conductance of the device (related to the inverse of the source-drain resistance) are best measured by considering the differential conductance with the derivative taken at constant $V_t$.

This method yields measurements at higher signal-to-noise ratio compared to digital method of taking derivatives. The differential conductance $G$ depends on top gate voltage $V_t$ or analyte concentration $m$ in solution as well as bias voltage $V_{ds}$. The quantity of interest is $\Delta G$, the change in conductance due either to a change in the top gate voltage $V_t$, or due to change in concentration $\Delta m$. A higher signal $\Delta G$ can be obtained in the region of negative $V_{ds}$ or positive $V_t$ for our nanosensor.

To demonstrate that the FET nanosensors are a robust platform for sensitive protein concentration determinations, we showed that there was a corresponding change in the differential conductance of the device that was dependent on the protein concentration of the analyte. As a demonstration of this principle, we used a monoclonal antibody association with its hapten, as a model for a molecular binding interaction. Typically, antibodies will bind selected epitopes with high specificities and affinity. The Antibiotin antibody (antibiotin) is known to specifically recognize biotin in either free or conjugated forms by many independent studies (Sigma-Aldrich, B3640).

A nanosensor device composed of 20 silicon wires in parallel (6 um long, 250 nm wide, 100 nm thick) was formed. The silicon nanowires were covered by 5 nm of an $Al_2O_3$ insulating layer as specified above, and then chemically modified by biotinamidocaproyl-labeled bovine serum albumin (BSA), providing a biotin-conjugate surface. Protein binding experiments were conducted in 1mM phosphate containing 1mM NaCl, pH=7.4.

To determine the protein binding contribution to conductance changes, solutions bearing different concentrations of the antibiotin are introduced via a syringe pump into the FET nanosensor chamber. Less than 2 minutes are required to register a stable reading of the conductance (Figure 4c). Changes in the differential conductance $G$ are observed consistent with different antibiotin concentrations as these solutions are pumped to the nanosensor surface (Figure 4a). The nanosensor accurately monitors antibiotin concentrations despite changes to order of additions of the antibiotin concentrations (Figure 4a and data not shown). Further, this data is delineated in plots of differential conductance as a function of protein concentration, showing that a typically shaped binding curve can be formed (Figure 4b). At higher concentrations of 0.5 microgram/ml, the device response saturates; presumably corresponding to the coating of the entire silicon nanowire surface. At low concentrations, the device response is rapid and precise.

Protein binding and concentration dependent measurements with the FET nanowire device are demonstrated by the antibiotin experiments diagrammed in Figure 4(a) and (b). Silicon nanochannels, functionalized with biotinylated bovine serum albumin (BSA), are used to detect antibiotin in 1 mM NaCl and 1 mM phosphate buffer solution. At the concentration of salt used in solution, the Debye screening length $\lambda_D$ at room temperature is about 9.6 nm. $\lambda_D$ is sufficiently large that the surface potential is sensitive to protein binding, but short enough to screen out biomolecules in solution (Stern *et al.*, 2007b). All solutions used in the measurements are dialyzed in order to keep constant ionic strength. The mechanism of ligand receptor binding can be simply explained by a two-state model. Assuming the dissociation constant of biotin-antibiotin binding is $K_{eq}$, then,

$$\text{Antibiotin} + \text{biotin} \rightarrow \text{Antibiotin-biotin-complex} \qquad (1)$$

The fractional occupancy of the binding sites is related to the biotin solution concentration *m* by a Langmuir adsorption isotherm. Assume the total bonding sites on the silicon wires are *N* and there are $a_p$ protein molecules attached on the surface:

$$a_p = \frac{N}{K_{eq}/m + 1} \quad (2)$$

By assuming the conductance change of the device is proportional to the charge concentration of the silicon nanowire surface, the data can be fit to the above two-state model, and a dissociation constant of $5.2 \times 10^{-10}$ M is calculated. This value compares favorably with an experimental $K_d$ value of $10^{-9}$ M reported in previous studies (Cui *et al.*, 2001b). Furthermore, the protein detection limit is 3 ng/ml (around 20 pM), when the signal-to-noise ratio is set to 1.

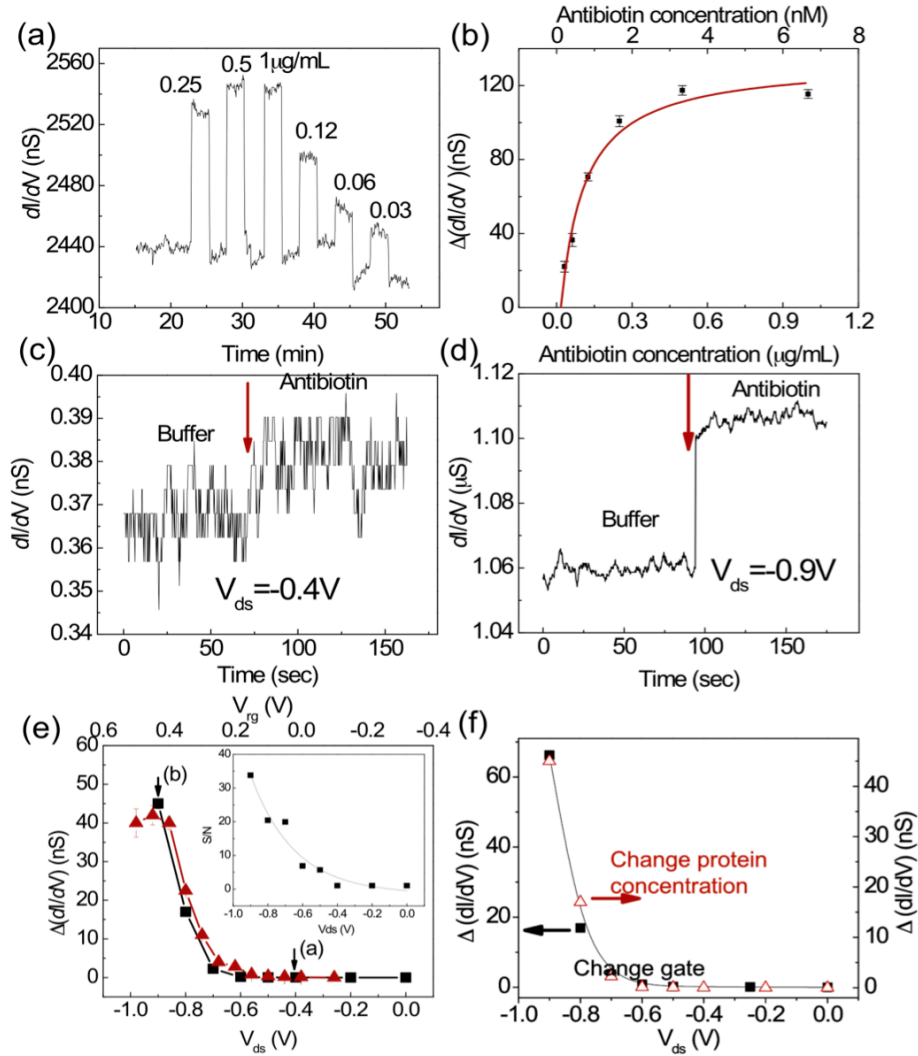

Figure 4. The FET nanosensor detects protein binding to the nanosensor surface. (a) Differential conductance versus concentration of antibiotin in phosphate buffer for a device composed of 20 wires of 250 nm width, $V_{ds}$ = −0.5 V, $V_{rg}$ = 0 V. (b) Change of conductance (with random error) versus antibiotin concentration. The solid line is fit by a two-state model. (c) (d) Device response to antibiotin at different bias $V_{ds}$=−0.4 V and −0.9 V (e) Device sensitivity change with bias and gate voltage; Inset, signal-to-noise ratio; (f) Comparison of conductance change introduced by 5 mV of reference gate voltage change (black dots, left axis) and 100 ng/ml of antibiotin solution (red triangles, right axis).

We also demonstrated that by adjusting the source drain bias voltage to the subthreshold region, the device sensitivity can be improved. The differential conductance change $\Delta g$ contributed by 100 ng/ml antibiotin is 0.02±0.01 nS at $V_{ds}$=−0.4 V (Figure 4c), while $\Delta g$ is 45±0.1 nS at $V_{ds}$=−0.9 V (Figure 4d). The signal-to-noise ratio increases from 2 to 34 (see Figure 4e inset). The signal due to 100 ng/ml antibiotin injection is plotted as a function of $V_{ds}$ and $V_{rg}$ (Figure 4e).

This plot clearly shows the effect of reverse bias amplification. The change $\Delta g$ above, due to concentration change at fixed reference gate voltage, can be compared to the change caused by varying the reference gate voltage while keeping the concentration fixed. Equivalence between the surface potential change and concentration change is then established. $\Delta g$ introduced by 100 ng/ml (660 pM) of the antibiotin monoclonal antibody is equivalent to a gate voltage change of 7.2 ± 0.3 mV (see Figure 4f).

These experiments illustrate the FET nanosensor capability to sensitively measure analytes in solution. Many applications are possible for this technology, because it combines high sensitivity with speed, reproducibility, and miniaturization.

### 5.3. Breast cancer tumor antigen determinations with FET nanosensors

Although image methods (mammograms) are routinely used in breast cancer treatment decisions, clinicians are seeking reliable early indicators from simpler and more cost-efficient procedures, such as blood tests. There are two major utilities for measuring the CA15.3 tumor antigens in the circulation of breast cancer patients. First, tests that yield the greatest level of sensitivity are of most importance for CA15.3 and early detection of the cancer. Current methods operate in at a baseline detection cutoff of >38U/ml. Therefore, tests that increase the sensitivity to lower CA15.3 levels are of great value. Secondly, additional measurements of CA15.3 are valuable in determining progression status for breast cancer patients on therapy. Here the CA15.3 biomarker is monitored as a prognostic marker as an indicator of late stage disease and/or metastasis. If CA15.3 levels rise, it is an indicator of progression, and the clinical team may redirect the treatment plan. CA15.3 in the 500-2500 U/ml range is monitored.

The specification of the analyte being detected on the nanosensor is determined by the tumor antigens that can be recognized. In the examples shown here, we selected tumor antigens where there are detection methods and reagents currently available as a proof-of-concept test.

Silicon nanowires were functionalized (Chen *et al.*, 2010) with the antibody CA15.3 specific to the soluble portion of the protein that is shed from the tumor and enters the blood stream. As discussed above, the conjugated selected antibody on the silicon nanowire FET device surface contributes to the desired specificity of the measurement. The silicon nanowire FET devices undergo a change in its electrical conductance (or differential conductance $G$) depending on the binding of CA15.3 to the respective antibody (Chen *et al.*, 2010). The ability of a nanoscale CA15.3 FET device to detect CA15.3 at the clinical relevant levels is shown in the following experiments.

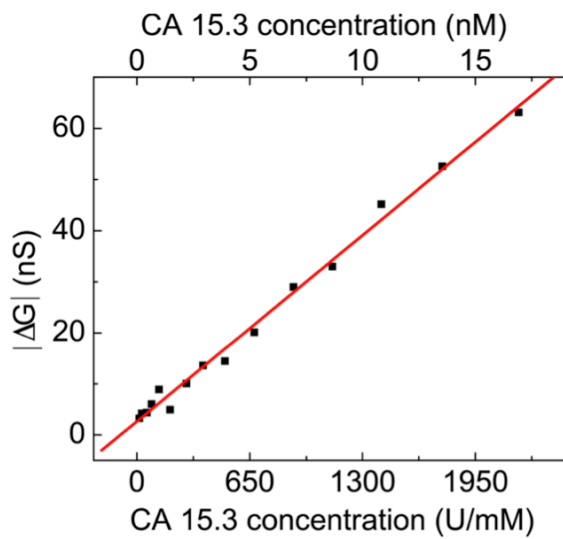

Figure 5. Measurements of CA15.3 binding events on FET biosensors functionalized with CA15.3 antibody.

The CA15.3 cancer antigen used for detection is dialyzed in solution with 150 µM NaCl and 10 µM phosphate buffer. Upon binding to antigen, the conductance of the nanosensor probe ("nanowire") changes (differential conductance $G$) in a CA15.3 concentration dependent manner (Figure 5). A measurement with an independently calibrated CA15.3 is also demonstrated. A real-time signal fluctuation is indicated when the syringe pump changes from buffer to 36U/ml CA15.3 in buffer. Furthermore, a plot of differential conductance versus concentration is a straight line at over three logarithms of concentrations range as shown.

*a) Detection sensitivity*

The detection limit of the conductance change is 1.68 nS, setting a limit in the detection of determined of 7.7 U/ml (~ 60 pM). The relationship of the differential conductance $G$ and CA15.3 concentration $m$ is shown in Figure 5. Over the range of clinical interest, the experimental data can be fit into a straight line from which we can derive a linear relationship between $G$ and $m$:

$$G(V_{ds}, V_{rg}, m) = G_1 + G_2 m \qquad (3)$$

Here, $G_1$ and $G_2$ are device dependent parameters and $G_1 = 307 \pm 2$ nS, $G_2 = -(28 \pm 1) \times 10^{-3}$ nS/(U/mL) from Figure 5. (At higher concentrations, the response is nonlinear and can be fit into an adsorption isotherm described above). With fixed CA15.3 concentration, differential conductance measurement can be done with varying reference gate voltage as shown in Figure 5:

$$G(V_{ds}, V_{rg}, m) = G_3 + G_4 V_{rg} \qquad (4)$$

where $G_3$ and $G_4$ are device dependent parameters and $G_3 = 304 \pm 2$ nS, $G_4 = 2.3 \pm 0.1$ nS/mV. For a given device, the parameter values serve to characterize the charge and equilibrium binding of a protein. An equivalence between the reference gate voltage $V_{rg}$ and the free CA 15.3 biomarker concentration $m$ is established from equation 3 and equation 4.

Additionally, the device response time is less than 60 seconds, indicating that the measurement is essentially instantaneous. This important feature of the FET nanosensor configuration may be primarily limited by the passive diffusion of the analyte to the nanosensor surface.

*b) Specificity of the nanosensor to CA15.3*

In addition, measurement of the unrelated monoclonal antibody, antibiotin, is used as a specificity control. Antibiotin, at differing protein concentrations below 20 nM shows no differential conductance $G$ different from baseline. The data indicate that the conductance change observed is primarily due to specific binding of the biomarker to its cognate antibody.

In summary, the nanosensor response was determined to operate as a function of CA15.3 concentration, demonstrated here over a span of 0 U/ml—2,300 U/ml, underscoring the wide dynamic range of this assay. Importantly, fluctuations in this dynamic range may be useful to a variety of clinical situations, for both early-stage diagnosis and later-stage prognosis, without reconfiguration.

When the control solution includes similar concentration of antibiotin, there is no detectable conductance change, which proves that the conductance change is due to selective binding between CA15.3 antibody and CA15.3 antigen. With a biotin concentration that is 30 times higher than the selective target (CA15.3 antigen), a specific conductance change can still be observed. The CA15.3 nanosensor responds as a function of CA15.3 concentration, demonstrated here over a span of 17 U/ml to 2,200 U/ml, and underscores the broad dynamic range that is achievable, which has utility for both early stage diagnosis (low concentration) and later-stage prognosis and monitoring to treatment (high concentration) without reconfiguration of the device and/or spectrum of nanosensors on a single chip.

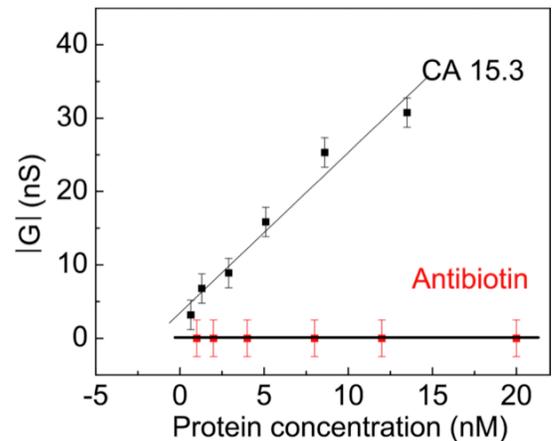

Figure 6. Selectivity of nanosensors based on antibody recognition. The differential conductance $G$ is determined from a CA15.3 nanosensor exposed to CA 15.3 or a protein control, antibiotin, at differing protein concentrations below 20nM shows no differential conductance $G$ different from baseline.

The fundamental advantage of the label-free nanosensor device architectures is the combination of high detection sensitivity and a standard semiconductor-based CMOS-compatible fabrication technique, suitable for scalable manufacturing. The unique FET configuration allows device-level signal amplification and the required degree of control to enable large-scale parallel architecture for detection of multiple target molecules. This combination is important for the screening, diagnosis and prognosis of diseases in a clinical setup, requiring fast analysis time, small sample volume and low cost. Further development of our array-based technique with high sensitivity protein assay is likely to be generally important for genomics and proteomics.

## 6. Discussion

The approach described here combines recent advances in the areas of nanoelectronics and biosensing. Eventual application of this work will be to enable highly parallel assay of protein biomarkers and signaling peptide molecules. In principle, the techniques can be used to detect any type of protein biomarkers, such as may occur in an air or liquid environment. Compared to the current assay methodologies, the proposed techniques facilitate a highly parallel approach for biomolecular recognition in small sample volumes. The key technology advantages – ultra high sensitivity, sensor flexibility, low sample volume, and highly scalable manufacturability – will facilitate point of care testing, and significantly reduce costs per test.

Recent advances in semiconductor technology, materials chemistry of surface conjugation, cancer diagnosis of tumor markers, and semiconductor device engineering are converging to produce greater diagnostic opportunities. In the semiconductor industry, the CMOS-compatible scalable architecture has enabled the creation and use of integrated chips with multi-million processing elements. Our nanoscale FET elements leverage the use of CMOS-compatible chip architecture (Stern *et al.*, 2007a) for ultrasensitive high-speed assay of multiple analytes. Although, for demonstration purposes, we have focused on the CA15.3 breast cancer biomarker, the device can be used for any tumor biomarker by the suitable adaptation of other specific antibodies on the device surface.

Others have also demonstrated cancer antigen detection at femtomolar sensitivities with devices fabricated by a "bottom up" method (Cui *et al.*, 2001a), using vapor–liquid–solid nanowire growth strategy. In this manufacturing method, silicon wires with feature widths between 5 nm to 30 nm are formed by modulating the nanocluster catalyst size on pre-patterned wafers containing electrodes. Just as with our nanosensors, these other approaches offer real-time, label-free and highly sensitive detection of a wide range of analytes, including proteins (Wang *et al.*, 2005), nucleic acids (Hahm & Lieber, 2004), small molecules (Chen *et al.*, 2008; Wang *et al.*, 2005; Wang *et al.*, 2008) and viruses (Patolsky *et al.*, 2004) in single-element or multiplex formats (Zheng *et al.*, 2005). However, the scalability of these methods are significantly impeded by the tedious chemically grown processes and assembly steps required with ultra-small diameters of the silicon wires. For industry applications, top-down manufacturing strategies such as ours, are much preferred.

There are several directions for improvements of CA15.3 testing that incorporate nanosensor strategies. First, it will be important to continue to adjust the assay sensitivity. The nanosensor platform depends on our ability to detect ultra-small currents in the nanowire in the sub-nanoampere range. The corresponding antigen detection sensitivity is estimated to be on the order of 1 U/microliter (one unit per microliter), which is a thousand-fold increase on the detection sensitivity of standard techniques such as chemiluminescent immunoassay. It should be noted that the current sensitivity is adequate for baseline determinations of the patients by comparison with the available clinical diagnostic tests. Not only does the small nanoscale size of the device contribute to the high sensitivity (as reflected in the corresponding change in the nanowire conductance), it also provides a small sensing area or binding site for the antigens. Since these antigens are large protein molecules (300-450 Kd), the change in conductance due to the binding of a single protein to the functionalized nanowire surface could be measured, providing the ultimate detection sensitivity. Refinements that investigate the transistor amplification of the detector sensitivity as a function of gate voltage promise to be an important development area. In addition, continuing a refinement of the surface dimensions towards ultrahigh sensitivity may also be of value.

Second, we expect that these nanosensors will be active over much broader dynamic ranges than was illustrated in this report. Since the FET response is a function of CA15.3 concentration, demonstrated here over a span of 0 U/ml—2,300 U/ml, underscores the wide dynamic range in the necessary clinically significant levels. Modifications that extend the dynamic range are clearly of interest for other settings, are actively being investigated. In cancer diagnostic applications, it may be favorable to utilize common nanosensor devices for CA15.3 and other tumor antigens, without device reconfiguration, but providing differential sensitivity ranges for different settings. These attributes may be preferred in the clinical context of providing early-stage diagnosis or late-stage prognosis and monitoring from a common platform.

Third, our nanosensor device response time is less than 60 seconds with a CA15.3 detection limit of ~1-2 U/ml (~10 pM). Experiments are underway to determine the factors driving the response time for antibody-antigen interactions, but a likely parameter is the diffusion process. Adjustments to the response times and sample processing functions are expected to be of assistance towards a point-of-care platform.

*Multiplexing with other breast cancer markers*

Multiplexed antibody arrays may be formed utilizing the same strategy, and also by using commercially available reagents. Additional breast cancer biomarkers may be valuable to examine in tandem with CA15.3. CA27.29 is also a mucin-associated antigen, which is detected by the monoclonal antibody B27.29, specific for the protein core of the MUC1 gene product. CA27.29 has been evaluated in comparison to other tumor markers including CA15.3 (Gion *et al.*, 1999). Specifically, in a detailed comparison of the diagnostic accuracy of CA27.29 and CA15.3, Gion et al (Gion *et al.*, 1999) show that CA27.29 discriminates primary breast cancer better than CA15.3, especially in patients with limited disease. In healthy subjects, CA15.3 was found to be significantly higher than CA27.29. In breast cancer patients, CA27.29 was found to be considerably higher than CA15.3, suggesting that CA27.29 is a better marker. The CA27.29 antigen is detected by monoclonal antibody B27.29, specific for the protein core of the MUC1 product. This antibody is already in use in fully automated versions of the CA27.29 assay such as the ACS:180 BR assay (Chiron Diagnostics) based on a conventional chemiluminescent immunoassay. In this technique, a mouse monoclonal antibody is incubated with both the patient sample and purified CA27.29 coupled covalently to paramagnetic particles in a solid phase. Both the solid-phase CA27.29 and the antigen in the sample compete for binding to the antibody. The resulting chemiluminescent signal is therefore found to be inversely proportional to the amount of antigen in the sample.

The American Society of Clinical Oncology (ASCO) recommended the use of a set of tumor markers for breast cancer in the 2007 clinical practice guideline (Harris *et al.*, 2007). In particular, CA15.3 and CA27.29 were evaluated for their ability to determine diagnosis and prognosis, monitor therapy, and predict recurrence of breast cancer after curative surgery and radiotherapy. CA27.29 is elevated in 29%, 36%, and 59% stages I, II and III cancers respectively. In comparison, the incidence of elevation for CA15.3 in the same samples was 15%, 23% and 54.5% for these patients, respectively (Gion *et al.*, 1999). Although a greater sensitivity towards earlier detection of disease progression may be evident from CA27.29 assays, it is clear that examining both markers in tandem may be beneficial. Thus, a technique capable of simultaneous detection of multiple biomarkers with high sensitivity in small volumes of samples is desirable.

Additional serum breast cancer markers, including CA125, CA19.9, and CEA, may also be productively examined. Further, circulating cell-bound markers may also be considered for eventual evaluation with nanosensors, provided that these assays achieve acceptable sensitivities and specificities. Investigations are underway with other cancer markers, such as HER2, EGFR, and other general epithelial cancer markers, in this regard. Advances in multiplexing with FET nanosensors are expected to allow device-level signal amplification and the required controllability to enable large-scale parallel processing architecture for detection of multiple target molecules.

The fundamental advantage of our label-free device architecture is the combination of high detection sensitivity and a standard semiconductor-based CMOS-compatible fabrication technique, suitable for scalable manufacturing. The combination of these nanosensor technology advances are important for the development of high quality screening, diagnosis and prognosis of diseases in the clinic, requiring fast analysis time, small sample volumes, and low cost. Further development of our array-based technique with high sensitivity protein assay may be broadly important to genomics and proteomics.

## 7. Future trends

It is attractive to consider these nanosensor platforms in light of future trends towards personalized medicine, because these technologies offer advantages of high sensitivity and speed at low cost.

In this chapter, we have discussed how nanochannel field effect transistor devices can be fabricated using CMOS-compatible lithography processes, familiar to the semiconductor industry. It is possible with nanochannel FET devices to build multiple sensor arrays integrated with current semiconductor technology. By addressing refinements to the design circuitry and the etching process, and multiplexing the functionalization steps, these device open promising avenues to detecting many analytes relevant to cancer diagnostics with high sensitivity.

a) *Multiplexing*

It is well recognized in cancer diagnostics and therapy selection that many tumor markers are present that may contribute to a better definition of the disease state. Yet, clinical access to these markers will often be limited by costs, as physicians do not want to order large numbers of individual tests. Thus, an avenue of great utility will be the ability to create a platform that provides highly multiplexed information, such as with the FET biosensor arrays depicted here. Implementing these FET biosensors with an appropriate circuit design provides the combined benefit of coordinating individual signal measurements as part of the management of highly multiplexed reads and high information content in tandem. Many innovations in the recording and interpretation of this multiplexed information for cancer diagnostics will follow.

The tumor antigen CA15.3 is typically reported as a single analyte test in clinical diagnostics practice. According to product labeling for different assays, these CA15-3 tests are indicated for the serial measurement of CA15-3 reactive antigenic determinants as an aid in the monitoring of patients previously diagnosed with breast cancer. CA15-3 measurements are approved for disease progression or response to therapy in conjunction with other clinical methods, such as imaging(Harris *et al.*, 2007). The CA15.3 assay can also be used as an aid in the detection of recurrence in previously treated Stage II and III breast cancer patients. Improvements to the CA15.3 measurements are likely to include the incorporation of additional serum markers in multi-marker algorithms. Further, changes to CA15.3 in response to chemotherapy treatment may provide patient benefit (Kim *et al.*, 2009). Tests with high sensitivity and specificity are required to make clinical decisions derived from biomarker levels during kinetic responses to chemotherapy.

In addition to utilities in breast cancer diagnostics, serum marker testing may also be beneficial in other cancer monitoring, such as for ovarian cancer. It has been demonstrated that multimarker algorithms of four or more serum markers (panel of CA125, HE4, CEA, and VCAM-1 in this example) outperform single prognostic markers (Yurkovetsky *et al.*). Also, it is evident that patient baseline level variations of many serum markers are an important constraint, and utilizing multiple independent markers in tandem is thought to aid a more robustly applicable test outcome.

*b) Miniaturization*
A striking benefit of these nanosensor technologies is the ability to implement them with dramatically reduced footprints. It is estimated that thousands of separate tests may be run in single chip devices with external dimensions of the chip of 1 cm$^2$.

*c) Point-of-care*
Device portability and reliability are the cornerstones of developing point-of-care diagnostics. Today, many clinical tests in the cancer diagnostics arena operate from highly sophisticated instrumentation and optics, and necessarily require highly skilled operators. Ostensibly, the use of nanosensors allows for the transition to a user interface of great simplicity. In operation, the devices of the future will have complex algorithms that read and process the individual biomarker outputs. However, these algorithms will function in the background of the device, making it possible to transform the diagnostic to a user-friendly interface that is clinically actionable.

## 8. References and Web Resources